\documentclass[
 reprint,
 amsmath,amssymb,
 aps,
]{revtex4-2}

\usepackage{graphicx}
\usepackage{dcolumn}
\usepackage{bm}
\begin{document}

\preprint{APS/123-QED}

\title{Chirality memory stored in magnetic domain walls in the ferromagnetic state of MnP}

\author{N. Jiang$^{1}$}
  \email{jiangnan@g.ecc.u-tokyo.ac.jp}
\author{Y. Nii$^{2,3}$}
\author{H. Arisawa$^{2}$}
\author{E. Saitoh$^{2,4,5,6}$}
\author{Y. Onose$^{2}$}

\affiliation{
$^{1}$Department of Basic Science, The University of Tokyo, Tokyo 153-8902, Japan.\\
$^{2}$Institute for Materials Research, Tohoku University, Sendai 980-8577, Japan.\\
$^{3}$PRESTO, Japan Science and Technology Agency (JST), Kawaguchi 332-0012, Japan.\\
$^{4}$Department of Applied Physics, The University of Tokyo, Tokyo 113-8656, Japan.\\
$^{5}$Advanced Science Research Center, Japan Atomic Energy Agency, Tokai 319-1195, Japan.\\
$^{6}$Advanced Institute for Materials Research, Tohoku University, Sendai 980-8577, Japan.
}

\date{\today}

\begin{abstract}
Chirality in a helimagnetic structure is determined by the sense of magnetic moment rotation. We found that the chiral information did not disappear even after the phase transition to the high-temperature ferromagnetic phase in a helimagnet MnP. The 2nd harmonic resistivity $\rho^{\rm 2f}$, which reflects the breaking down of mirror symmetry, was found to be almost unchanged after heating the sample above the ferromagnetic transition temperature and cooling it back to the helimagnetic state. The application of a magnetic field along the easy axis in the ferromagnetic state quenched the chirality-induced $\rho^{\rm 2f}$. This indicates that the chirality memory effect originated from the ferromagnetic domain walls.
\end{abstract}

\maketitle

Chirality is a lack of mirror symmetry in matter. If a molecule has chirality, its mirror image is different from the original structure. In other words, the molecule has two distinguishable states (enantiomer) with different chiralities. In the human body, some molecules, such as amino acids, are homo-chiral, meaning that they show only one certain chiral state. In biology, it is important to elucidate how chiral information is transferred and memorized.\cite{chiral1,chiral2}.

Chirality also appears in magnetic structures. One example is a helical magnetic structure, in which the ordered direction of the magnetic moment spatially rotates in a plane perpendicular to the propagation vector (Fig. 1 (a)). The sense of rotation is reversed by any mirror operation and therefore determines the chiral state. Similar magnetic chirality shows up in a Bloch-type domain wall (DW) in ferromagnets (Fig. 1 (a)). The magnetic moment rotates on the domain wall, and the sense of rotation is responsible for the chirality, similarly to helimagnets. In this paper, we report the transfer of chiral information from a helical magnetic structure to a ferromagnetic DW during a helimagnetic to ferromagnetic phase transition in an itinerant helimagnet MnP.

\begin{figure*}
\includegraphics[width=12cm]{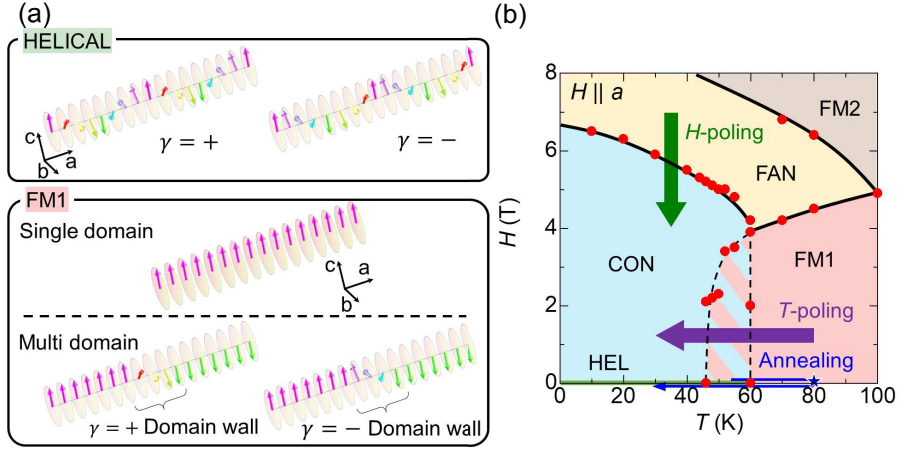}
\caption{\label{fig:wide}(a) Illustrations of helical magnetic structures and Bloch-type ferromagnetic domain walls. $\gamma =\pm$ denotes the chirality. (b) The phase diagram of a micro-fabricated MnP sample for $H || a$. The dots are the phase boundaries estimated by the magnetic field dependence of the electrical resistivity\cite{supple}. The dashed and solid lines are guides for the eyes. In the hatched region, the realized magnetic structure depends on the hysteresis. The arrows illustrate the $H$-poling, $T$-poling and annealing procedures (see main text).}
\end{figure*}

\begin{figure}
\includegraphics[width=7cm]{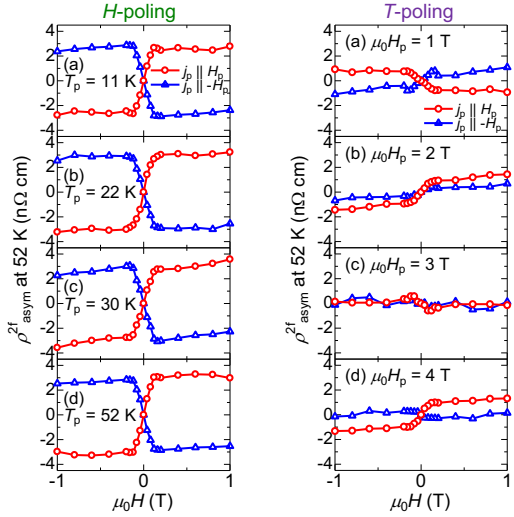}
\caption{(a)-(d) Magnetic field dependence of $\rho^{\rm 2f}_{\rm asym} = (\rho^{\rm 2f} (H$) $-$ $\rho^{\rm 2f} (-H$))/2 at 52 K after the $H$-poling procedure performed at (a) $T_{\rm p}=$ 11 K, (b) 22 K, (c) 30 K and (d) 52 K with dc electric currents $j_{\rm p}$ parallel to and anti-parallel to magnetic fields $H_{\rm p}$. The magnitude of $j_{\rm p}$ was $4.2 \times 10^{8}$ $\rm Am^{-2}$ at 11 K and $8.5 \times 10^{8}$ $\rm Am^{-2}$ at 22 K, 30 K and 52 K. The magnitude of ac electric current for measuring the 2nd harmonic resistivity $j_{\rm ac}$ was $5.9 \times 10^{8}$ $\rm Am^{-2}$. (e)-(h) Magnetic field dependence of $\rho^{\rm 2f}_{\rm asym}$ at 52 K after the $T$-poling procedure at (e) $\mu_{0}H_{\rm p}$ = 1 T, (f) 2 T, (g) 3 T and (h) 4 T with $j_{\rm p} \parallel H_{\rm p}$ and $j_{\rm p} \parallel -H_{\rm p}$. The magnitudes of $j_{\rm p}$ and $j_{\rm ac}$ were $8.5 \times 10^{8}$ $\rm Am^{-2}$ and $5.9 \times 10^{8}$ $\rm Am^{-2}$, respectively.} 
\end{figure}

\begin{figure*}
\includegraphics[width=13cm]{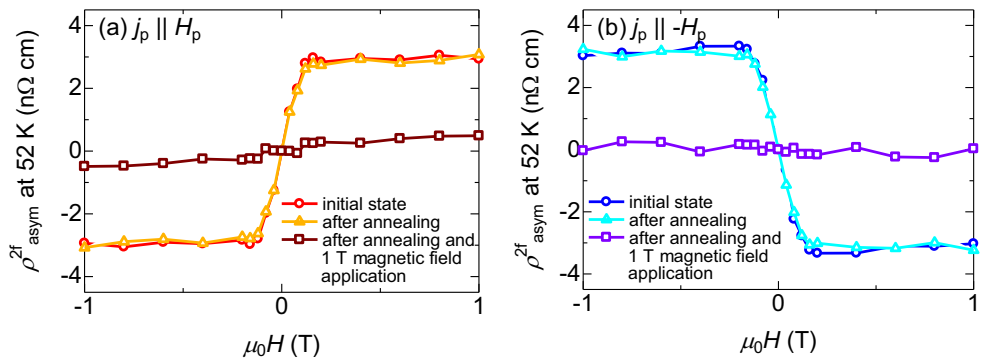}
\caption{\label{fig:wide}(a), (b) Magnetic field dependence of $\rho^{\rm 2f}_{\rm asym}$ at 52 K at the initial state (just after H-poling, circles), that after annealing at 80 K (triangles), and that after annealing at 80 K and application of 1 T magnetic field along the $c$-axis (squares). The initial $\rho^{\rm 2f}_{\rm asym}$ was positive ($j_{\rm p} \parallel H_{\rm p}$) and negative ($j_{\rm p} \parallel -H_{\rm p}$) in (a) and (b), respectively.} 
\end{figure*}

\begin{figure}
\includegraphics[width=6cm]{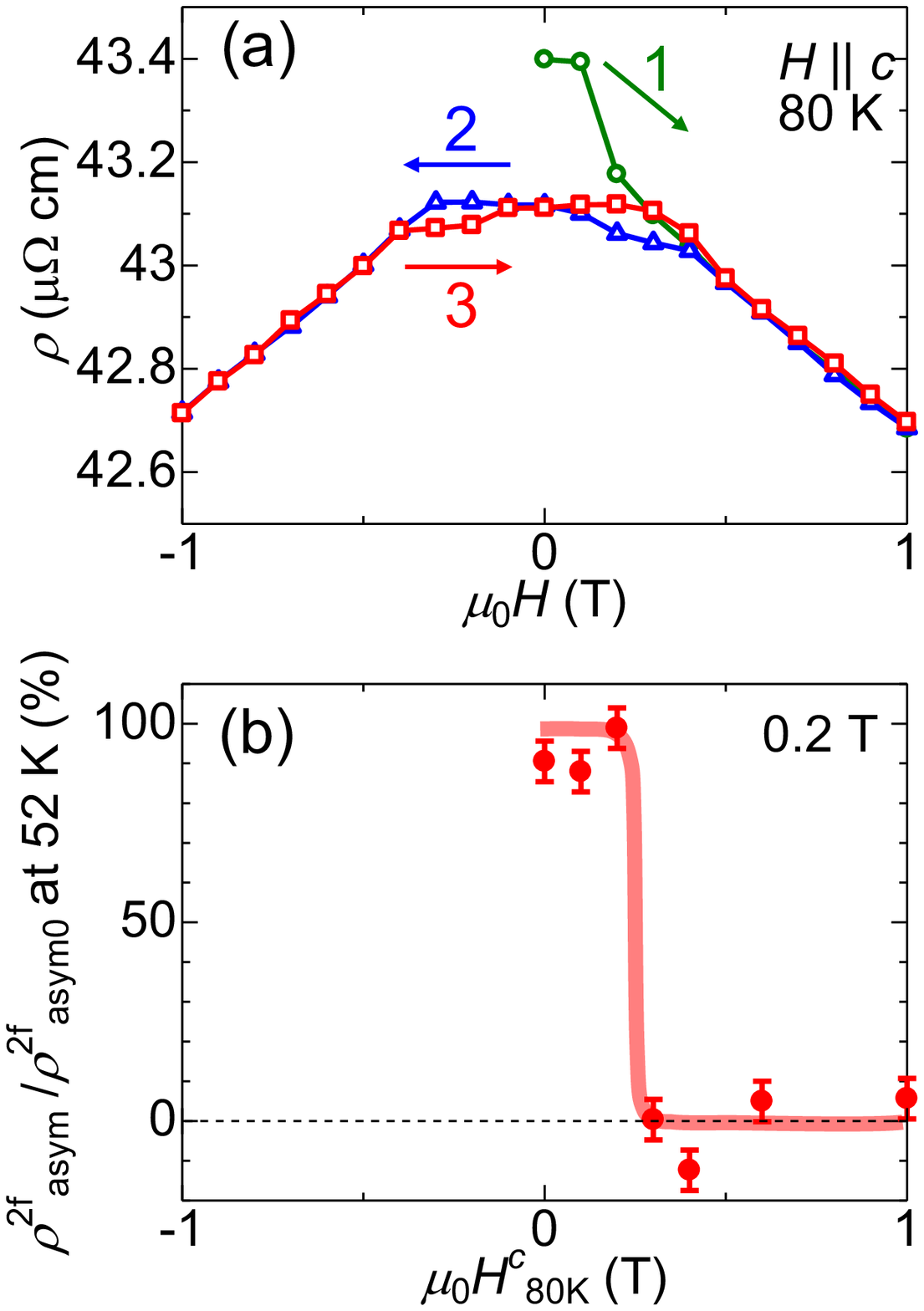}
\caption{(a) Magnetic field dependence of the resistivity $\rho$ at 80 K. The magnetic field was parallel to the magnetic easy axis ($c$-axis). Circles represent the data measured just after a phase transition from the helical to the ferromagnetic phase at 0 T. Triangles and squares represent the resistivity measured while decreasing the magnetic field from 1 T to -1 T and while increasing the magnetic field from -1 T to 1 T, respectively. (b) $\rho^{\rm 2f}_{\rm asym}$ at 0.2 T measured after annealing and application of magnetic field $H^{c}_{\rm 80K}$ normalized by that before the annealing process ($\rho^{\rm 2f}_{\rm asym0}$) plotted against $\mu_{0}H^{c}_{\rm 80K}$. The magnetic field $H^{c}_{\rm 80K}$ was applied along the $c$-axis at 80 K. The error bar is the standard deviation of $\rho^{\rm 2f}_{\rm asym}/\rho^{\rm 2f}_{\rm asym0}$ for the $\mu_{0}H^{c}_{\rm 80K}$ = 1 T data (Fig. 3 (b) squares), which is assumed to vanish.} 
\end{figure}

Fig. 1(b) shows the phase diagram of MnP\cite{MnP1,MnP2,MnP3,MnP4,MnP5,MnP6,MnP7,MnPphase,MnP8}, which was constructed based on magnetoresistance measurements of a microscale sample ($10\times 20\times 1$ $\rm \mu m^{3}$) fabricated for the present study by using the focused ion beam technique \cite{supple}.
It is quite similar to that of a bulk sample reported in a previous study\cite{MnPphase}. The ferromagnetically ordered phase (FM1) was stable even at zero magnetic field above 60 K, and the magnetic moments aligned along the $c$-axis. In this magnetic state, Bloch-type magnetic domain walls were observed by Lorentz transmission electron microscopy\cite{MnP8}.
On the other hand, in the helimagnetic state (HEL), which showed up below 44 K, the helical plane was normal to the propagation vector along the $a$-axis\cite{MnP4}. The intermediate hatched region indicates the metastable region typical of the first-order phase transition. When the magnetic field was applied parallel to the $a$-axis in the helimagnetic state, the magnetic moments were tilted, forming a conical magnetic state (CON). As the magnetic field increased further before the magnetic moments became completely aligned along the $a$-axis (FM2), there emerged a fan structure (FAN), in which the magnetic moments were within the $ac$ plane, and the angle between the magnetic moment and the propagation vector spatially oscillated along the propagation vector. In our previous work\cite{jiang}, we controlled the chirality of the helimagnetic structure by the simultaneous application of a magnetic field and a dc electric current and probed the controlled chirality by the nonreciprocal electronic transport, which is the field asymmetric component of 2nd harmonic resistivity\cite{EMC0,EMC1,EMC2,EMC3}. The controlled chirality was shown to be dependent on whether the applied magnetic field and the dc electric current are parallel or antiparallel to each other. In this paper, we show how the controlled chirality in the helimagnetic state was preserved or disappeared in the high-temperature ferromagnetic state.

To reveal how the chirality information is affected by external stimuli in the course of phase transitions, we show the field-asymmetric 2nd harmonic resistivity after the simultaneous application of a magnetic field and a dc electric current in the course of FAN-to-CON and FM1-to-CON phase transitions.
For the FAN-to-CON transition, we applied a dc electric current $j_{\rm p}$ parallel or antiparallel to the magnetic field $H_{\rm p}$ along the propagation vector ($a$-axis) when the FAN-to-CON transition field was traversed. The magnitude of $\mu_{\rm 0}H_{\rm p}$ was slowly decreased from 7 T to 4 T at a rate of 0.6 $\rm Tmin^{-1}$. Then, we removed the dc electric current $j_{\rm p}$. We call the series of these processes the \lq\lq $H$-poling procedure\rq\rq (Fig. 1 (b)).
In our previous paper\cite{jiang}, we performed the $H$-poling procedure at 51 K and measured the 2nd harmonic resistivity to show the chirality control. Here we performed the $H$-poling procedure at various temperatures and measured the 2nd harmonic resistivity at the same temperature, 52 K. Figures 2 (a)-(d) show the asymmetric component of the 2nd harmonic resistivity $\rho^{\rm 2f}_{\rm asym} = (\rho^{\rm 2f} (H$) $-$ $\rho^{\rm 2f} (-H$))/2 at 52 K after the $H$-poling at various temperatures $T_{\rm p}$. Here $\rho^{\rm 2f}$ is the observed 2nd harmonic resistivity. As discussed in the previous paper\cite{jiang}, $\rho^{\rm 2f}_{\rm asym}$ is sensitive to the breaking of chiral symmetry, whereas the symmetric component seems to be due to sample nonuniformity and/or electrode contact. Notable $\rho^{\rm 2f}_{\rm asym}$ was discerned for all the poling conditions. The signal was reversed by the reversal of the dc poling electric current. These features indicate that the chirality is effectively controlled by the application of the electric current and the magnetic field in the course of the FAN-CON transition, irrespective of temperature.  

While the FAN-CON transition is a second-order phase transition, the FM1-CON transition is a first order one. For this reason, the chirality responses in the course of these phase transitions were quite different. We tried to control the chirality by a similar poling procedure via the FM1-CON transition (Fig. 1 (b)).
We applied $j_{\rm p}$ parallel or antiparallel to $H_{\rm p}$ along the $a$-axis when the transition temperature was traversed. The temperature was slowly decreased from 80 K to 30 K at a rate of 0.5 $\rm Kmin^{-1}$ with a constant magnitude of $H_{\rm p}$. Then, we removed $j_{\rm p}$. We call the series of these processes the \lq\lq $T$-poling procedure\rq\rq.
Figures 2 (e)-(h) show $\rho^{\rm 2f}_{\rm asym}$ at 52 K after $T$-poling procedures at various $H_{\rm p}$. The magnitude was much smaller than the case of the $H$-poling. It should be noted that the signal was not reproducible\cite{supple}. These features indicate that the $T$-poling is not effective for chirality control.

Thus, the chirality was insensitive to the external stimuli when traversing the first-order FM1-CON transition. We also found that the chiral information was preserved through this phase transition at zero magnetic field, as shown below.
In Figs. 3 (a) and (b), we show $\rho^{\rm 2f}_{\rm asym}$ just after the $H$-poling and after the $H$-poling and the subsequent annealing process (heating up to 80 K, cooing down to 30 K, and going back to the original measuring temperature of 52 K, as shown in Fig. 1 (b)). The $\rho^{\rm 2f}_{\rm asym}$ signal was almost unchanged even after the annealing process, irrespective of the sign of $\rho^{\rm 2f}_{\rm asym}$. This observation suggests that the chiral information is preserved even in the FM1 phase at zero field. Then, the question is how the chiral information is preserved in the collinear ferromagnet. This can be answered by considering the effect of magnetic field application along the magnetic easy axis ($c$-axis) at 80 K. 
After heating the sample up to 80 K, we rotated the sample device in the superconducting magnet and applied a magnetic field $\mu_{0}H^{c}_{\rm 80K}$ as large as 1 T along the $c$-axis. Then, we decreased the temperature to 30 K at 0 T and increased it again to the original measuring temperature of 52 K.  
Figures 3 (a) and (b) also show $\rho^{\rm 2f}_{\rm asym}$ at 52 K after this annealing process with the application of a magnetic field $H^{c}_{\rm 80K}$. It is clear from this figure that the $\rho^{\rm 2f}_{\rm asym}$ signal was quenched by the application of $H^{c}_{\rm 80K}$. This result implies that it is the ferromagnetic domain walls that memorize the chiral information.

To confirm the chirality memory effect of the ferromagnetic domain walls, we investigated the $H^{c}_{\rm 80K}$ dependence of $\rho^{\rm 2f}_{\rm asym}$. We measured $\rho^{\rm 2f}_{\rm asym}$ just after $H$-poling and after $H$-poling and the subsequent annealing process with the application of various magnetic fields $H^{c}_{\rm 80K}$. In Fig. 4(b), we plot $\rho^{\rm 2f}_{\rm asym}$ measured after the annealing process normalized by that before the annealing process ($\rho^{\rm 2f}_{\rm asym
0}$) as a function of $H^{c}_{\rm 80K}$, compared with the linear resistivity in magnetic fields parallel to the $c$-axis. The linear resistivity showed hysteretic behavior. The initial state of the measurement was zero magnetic field just after heating from the helimagnetic state. The resistivity was relatively large at the initial state and decreased with the magnetic field (circles). After the application of a large magnetic field, the resistivity showed a butterfly type hysteresis loop (triangles and squares) and was smaller than the initial resistivity. These features indicate that the magnetic domain walls present at the initial state increased the resistivity, which disappeared above 0.4 T. Importantly, $\rho^{\rm 2f}_{\rm asym}/\rho^{\rm 2f}_{\rm asym0}$ was quenched at almost the same magnetic field as shown in Fig. 4 (b). This strongly supports the scenario that the magnetic domain walls memorized the chiral information in the FM1 phase.   

In conclusion, we have demonstrated a chirality memory effect of ferromagnetic domain walls. The chirality of the helical magnetic structure is preserved even after annealing in the high-temperature ferromagnetic state at zero field.  
The origin of this memory effect is ascribed to the Bloch-type domain walls in the ferromagnetic state. Previously, similar memory effects have been observed in multiferroic helimagnets\cite{mnwo4,co4nb2o9,cuo,gddymno2,cufegao1,cufegao2,gddymno1,cufeo2}. Some of these effects are suggested to be caused by the so-called polar nanoregion\cite{mnwo4,co4nb2o9,cuo,gddymno2}. In other cases, the charge degree of freedom seems to be responsible for the memory effect\cite{cufegao1,cufegao2,gddymno1}. Beilsten-Edmands $et$ $al$. speculated, based on Monte Carlo simulation, that the antiferromagentic domain wall is responsible for the memory effect in the case of $\rm CuFeO_{\rm 2}$\cite{cufeo2}. The present result unambiguously shows, based on the experimental result of magnetic field application along the easy axis, that the Bloch-type domain walls are responsible for the memory effect in the case of MnP. This suggests that the magnetic domain walls work as small chiral nuclei in the nonchiral phase, which are important in determining the chirality after the first order nonchiral-chiral phase transition. 

The authors thank G.E.W. Bauer, Y. Shimamoto, Y. Togawa, and J. Ohe for fruitful discussions. The crystal growth was carried out by joint research in the Institute for Solid State Physics, The University of Tokyo with the help of R. Ishii and Z. Hiroi. The fabrication of the sample device was carried out partly by collaborative research in the Cooperative Research and Development Center for Advanced Materials, Institute of Materials Research, Tohoku University with the help of K. Takanashi and T. Seki. This work was supported in part by JSPS KAKENHI grant numbers JP16H04008, JP17H05176, JP18K13494, JP19H05600, and JP20K03828, the JST ERATO Spin Quantum Rectification Project (JPMJER1402), PRESTO grant number JPMJPR19L6, and the Mitsubishi foundation. N.J. is supported by a JSPS fellowship (No. JP19J11151).


%

\end{document}


\preprint{APS/123-QED}

\title{Supplemental Material \\Chirality memory stored in magnetic domain walls in the ferromagnetic state of MnP}

\author{N. Jiang$^{1}$}
  \email{jiangnan@g.ecc.u-tokyo.ac.jp}
\author{Y. Nii$^{2,3}$}
\author{H. Arisawa$^{2}$}
\author{E. Saitoh$^{2,4,5,6}$}
\author{Y. Onose$^{2}$}

\affiliation{
$^{1}$Department of Basic Science, The University of Tokyo, Tokyo 153-8902, Japan.\\
$^{2}$Institute for Materials Research, Tohoku University, Sendai 980-8577, Japan.\\
$^{3}$PRESTO, Japan Science and Technology Agency (JST), Kawaguchi 332-0012, Japan.\\
$^{4}$Department of Applied Physics, The University of Tokyo, Tokyo 113-8656, Japan.\\
$^{5}$Advanced Science Research Center, Japan Atomic Energy Agency, Tokai 319-1195, Japan.\\
$^{6}$Advanced Institute for Materials Research, Tohoku University, Sendai 980-8577, Japan.
}

\maketitle
\clearpage
\newpage
\section{An image of scanning electron microscopy, and temperature and magnetic dependences of linear resistivity for the present MnP sample.}

\begin{figure}[ht]
\includegraphics[width=12cm]{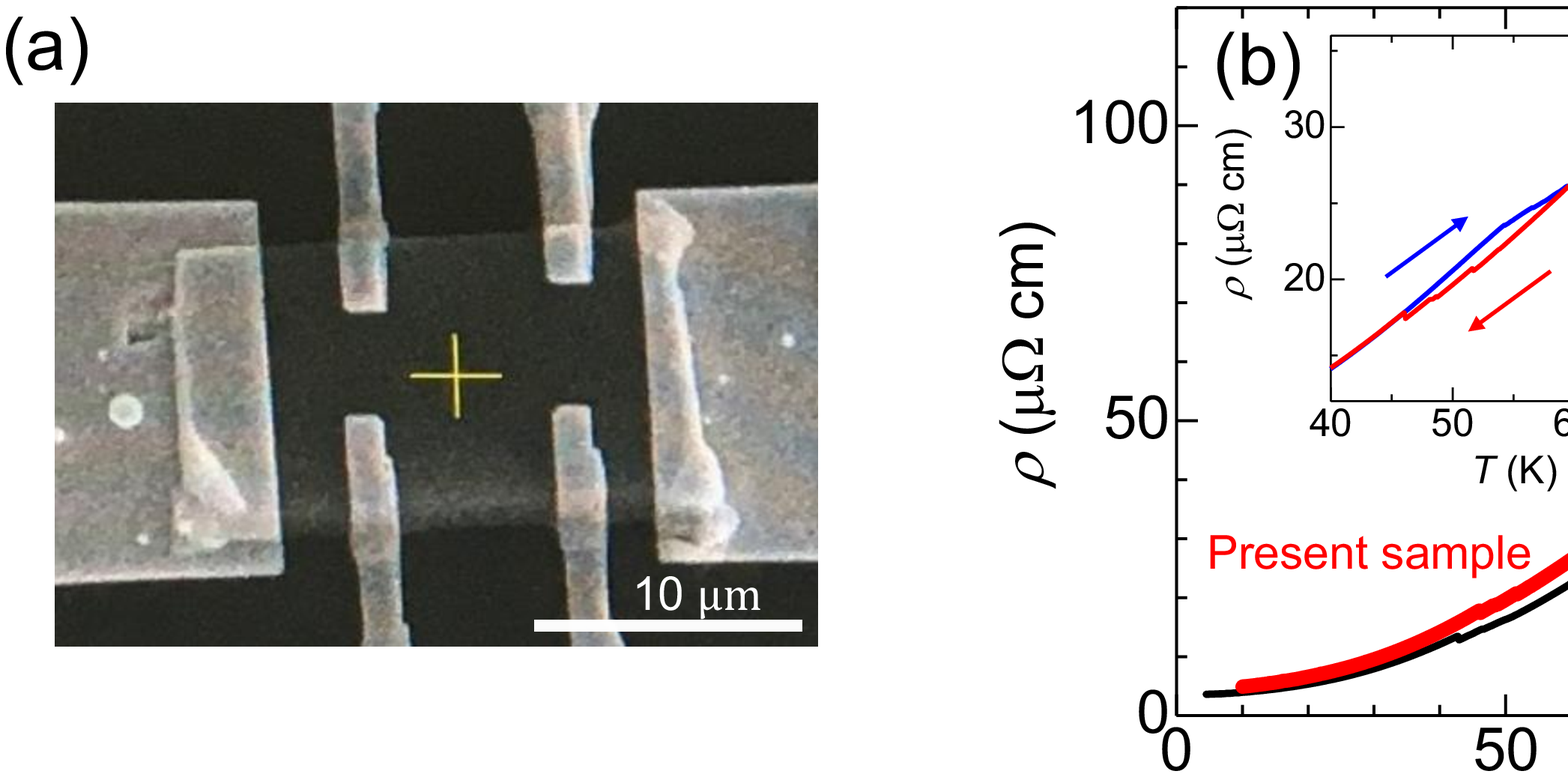}
\caption{(a) Scanning electron microscopy image of the micro-fabricated MnP sample. (b) Temperature dependence of the resistivity for the present MnP sample fabricated by using a focused ion beam. The resistivity of our previous sample is reproduced for comparison\cite{jiang}. The inset shows the history dependence of resistivity around the helical--ferromagnetic phase boundary.}
\end{figure}

\begin{figure}[ht]
\includegraphics[width=10cm]{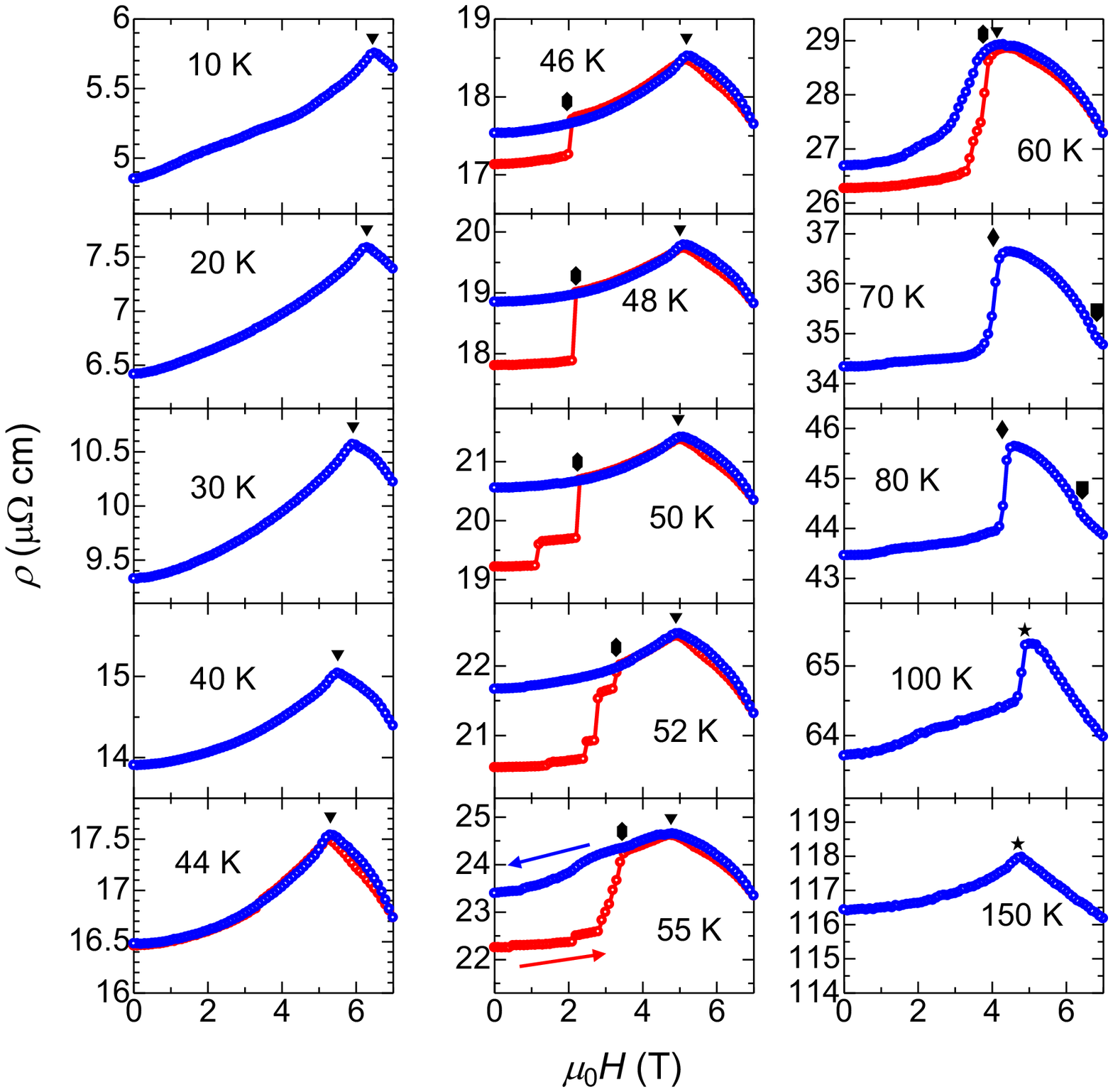}
\caption{Magnetic field dependence of the linear electrical resistivity at various temperatures. The hystereses were measured between 44 K and 60 K. Before the measurement in this temperature region, the temperature was increased to 80 K and then decreased to the measured temperature in the absence of a magnetic field. The red data shows the resistivity measured in the field-increasing process just after the zero-field cooling. The blue curves are the data for the field-decreasing process. Symbols indicate the metamagnetic transition fields shown in Fig. 1 (b) in the main text. Triangles, hexagons, pentagons, rhombuses, and stars represent FAN-to-CON, FM1-to-CON, FM2-to-FAN, FAN-to-FM1, and FM2-to-FM1 phase transitions, respectively. All the data are quite similar to those shown in the supplemental material of ref. 1.}
\end{figure}

\clearpage
\newpage

\section{$\rho^{\rm 2f}_{\rm asym}$ signal after various $H$-poling procedures.}

\begin{figure}[ht]
\includegraphics[width=11cm]{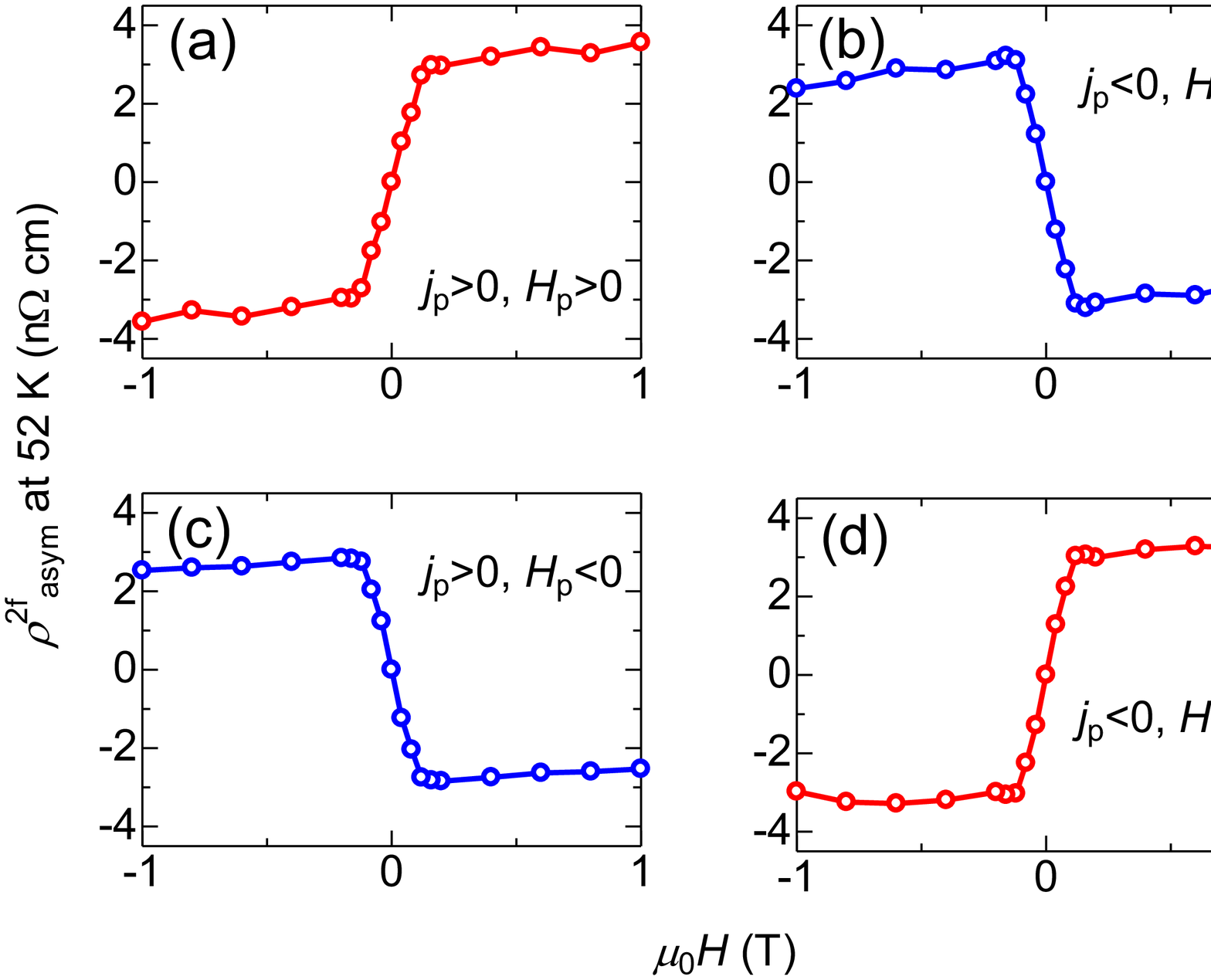}
\caption{(a)-(d) Magnetic field dependence of $\rho^{\rm 2f}_{\rm asym}$ after the $H$-poling procedure at 52 K with the positive and negative magnetic fields $H_{\rm p}$ and dc electric currents $j_{\rm p}$. The magnitudes of $j_{\rm p}$ and ac electric current for the measurement $j_{\rm ac}$ were $8.5 \times 10^{8}$ $\rm Am^{-2}$ and $5.9 \times 10^{8}$ $\rm Am^{-2}$, respectively. The sign of $\rho^{\rm 2f}_{\rm asym}$ depended on whether $H_{\rm p}$ and $j_{\rm p}$ were parallel or antiparallel, which is consistent with our previous paper.
While we only show the result of two conditions in main text, all the four conditions are shown here. }
\end{figure}

\newpage

\section{Lack of reproducibility for the $\rho^{\rm 2f}_{\rm asym}$ signal just after the $T$-poling.}

We found that the $\rho^{\rm 2f}_{\rm asym}$ signal after the $T$-poling was not reproducible. In Fig. 4, we show three observed signals of $\rho^{\rm 2f}_{\rm asym}$ measured with the same condition; they were observed just after the $T$-poling procedure at $\mu_{0}H_{\rm p}$ = 1 T for $j_{\rm p} \parallel -H_{\rm p}$. While 1st and 2nd data are similar to each other, the 3rd data is quite different from the other two.

\begin{figure}[ht]
\includegraphics[width=7cm]{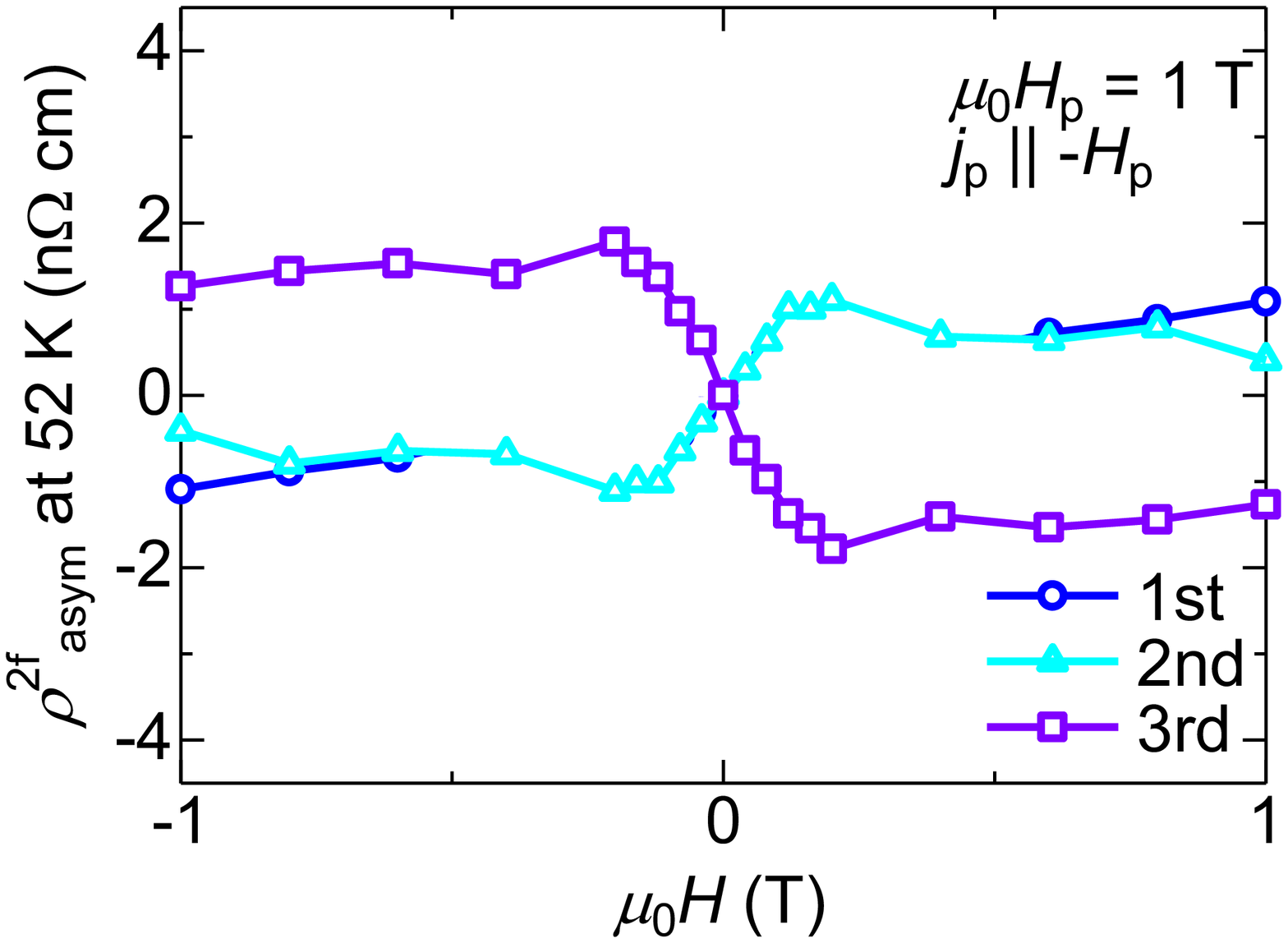}
\caption{Three $\rho^{\rm 2f}_{\rm asym}$ data measured with the same condition. They were observed just after the $T$-poling procedure with $j_{\rm p} \parallel -H_{\rm p}$. The magnitudes of $\mu_{0}H_{\rm p}$ and $j_{\rm ac}$ were 1 T and $5.9 \times 10^{8}$ $\rm Am^{-2}$, respectively.}
\end{figure}


%